\documentclass[5p,number]{elsarticle}

\usepackage{lineno,hyperref}
\usepackage{graphicx}
\usepackage{setspace}
\usepackage{lineno}
\usepackage{amsmath}
\usepackage{amsmath}
\biboptions{sort&compress}
\usepackage{pdfpages}

\journal{Journal of \LaTeX\ Templates}

\begin{document}

\begin{frontmatter}

\title{MuSIC@Indiana: an effective tool for accurate measurement of fusion with low-intensity radioactive beams} 

\author[chem,ceem]{J. E. Johnstone}
\author[chem,ceem]{Rohit Kumar}
\author[chem,ceem]{S. Hudan}
\author[chem,ceem]{Varinderjit Singh}
\author[chem,ceem]{R. T. deSouza \corref{cor1}}
\author[ND]{J. Allen}
\author[ND]{D.~W. Bardayan}
\author[ND]{D. Blankstein}
\author[ND]{C. Boomershine}
\author[ND]{S. Carmichael}
\author[ND]{A.~M. Clark}
\author[ND]{S. Coil}
\author[ND]{S.~L. Henderson}
\author[ND]{P.~D. O'Malley}

\cortext[cor1]{Corresponding author email address: desouza@indiana.edu}

\address[chem]{Department of Chemistry, Indiana University, 800 E. Kirkwood Ave., Bloomington, Indiana 47405, USA}
\address[ceem]{Center for Exploration of Energy and Matter, Indiana University, 2401 Milo B. Sampson Lane, Bloomington, Indiana 47408, USA}
\address[ND]{Department of Physics, University of Notre Dame, Notre Dame, Indiana 46556, USA} 

\begin{abstract}
The design, construction, and characterization of the Multi-Sampling
Ionization Chamber, MuSIC@Indiana, are described. This detector provides efficient 
and accurate measurement of the
fusion cross-section at near-barrier energies. The response of the detector to low-intensity beams of $^{17,18}$O, $^{19}$F, $^{23}$Na, $^{24,26}$Mg, $^{27}$Al, and $^{28}$Si at E$_{lab}$ = 50-60 MeV was examined. MuSIC@Indiana was commissioned by 
measuring the $^{18}$O+$^{12}$C fusion excitation function for 11 $<$ E$_{cm}$ $<$ 20 MeV using CH$_{4}$ gas.  A simple, effective analysis cleanly distinguishes
proton capture and two-body scattering  events from fusion on carbon. 
With MuSIC@Indiana, measurement of 15 points on the excitation function for a single incident beam energy is achieved.
The resulting excitation function is shown to be in good agreement
with literature data.
\end{abstract}


\end{frontmatter}


\section{Introduction}
\label{section:Intro}
The structure and reactions of neutron-rich isotopes is presently a topic of significant interest \cite{NSACPlan2015}. As nuclei become more neutron-rich their properties are expected to change and new collective modes may emerge.
The availability of neutron-rich beams at radioactive
beam facilities now allows the systematic exploration 
of fusion for an isotopic chain of neutron-rich nuclei \cite{Carnelli14,Singh17,Vadas18,Hudan20,deSouza21}.
While the next generation of radioactive beam facilities, 
such as the Facility for Rare
Isotope Beams (FRIB), will provide radioactive beams closer to the neutron 
drip-line than ever before \cite{FRIBUpgrade}, it also presents experimental challenges.
Due to their short half-lives, these exotic beams will 
only be available at low intensity mandating use of an effective and efficient means for accurately measuring fusion probability.

The low intensity of exotic radioactive beams suggests that a thick target approach
should be used. Thick target approaches have previously been used in the measurement of
fusion by identifying the fusion products via their characteristic $\gamma$-radiation as they de-excite \cite{Thomas85}. However, utilizing this approach
requires accurate knowledge of the $\gamma$ detection efficiency -- which is often low -- as well as
knowledge of the decay properties of the neutron-rich fusion products -- which may not exist.  

An alternative approach is to use an active target in which direct detection of the primary charged fusion products provides the signal that fusion
has occurred. 
A Multi-Sampling Ionization Chamber (MuSIC) detector \cite{Carnelli14} provides an
effective means of measuring the fusion cross-section by identifying the heavy fusion product. 
While MuSIC detectors were originally developed for use in high-energy heavy-ion experiments \cite{Christie87,Kimura90,Boyd92}, more 
recently their use has been extended 
to low energy nuclear reactions namely the 
measurement of the fusion excitation 
function for $^{10-15}$C+$^{12}$C 
\cite{Carnelli14,Carnelli15} and
$^{17}$F+$^{12}$C \cite{Asher21}, or studies of ($\alpha$,n)/($\alpha$,p) reactions 
\cite{Avila16,Avila17}.

The MuSIC approach provides a couple of intrinsic advantages over the typical thin-target measurement. Traditional thin-target
measurements were performed with limited angular coverage, identifying the fusion products by either $\Delta$E-E \cite{Eyal76, Kovar79} or ETOF \cite{Steinbach14a} techniques. Extraction of the fusion cross-section thus required integration of the angle and energy distributions for the individual heavy product introducing an uncertainty into the total extracted fusion cross-section. Use of a MuSIC detector provides a direct integrated measure of the fusion cross-section. 
In contrast
to the thin-target approach where the incident beam energy must be changed, 
MuSIC detectors allow measurement of multiple points on the excitation function simultaneously \cite{Carnelli15}.
In addition, MuSIC detectors are 
self-normalizing since the incident beam is detected by the same detector as the reaction products.
These advantages make MuSIC detectors an 
efficient means for measuring fusion excitation functions 
for neutron-rich nuclei when available beam intensities 
are limited.

This paper describes the design and construction 
of a MuSIC detector at Indiana University designated MuSIC@Indiana,
along with its characterization both with an $\alpha$-source and $^{18}$O beam. To commission the detector, the fusion excitation function for $^{18}$O + $^{12}$C was measured. A simple analysis of the data is described which allows one to isolate fusion by distinguishing it from events corresponding to proton capture and two-body scattering.
The measured excitation function is compared with previously reported fusion excitation functions for this reaction in the literature.

\section{Detector design and characterization}
\label{section:Design}

A MuSIC detector consists of a transverse-field, Frisch-gridded ionization chamber with the anode subdivided into strips along the beam direction. The signal from each anode segment is readout independently allowing the energy deposit of an ionizing particle to be sampled. As it traverses the detector, the beam loses energy in the detector gas at a rate characterized by its specific ionization. If a fusion event occurs in the detector the compound nucleus formed is higher in atomic and mass number than the incoming beam. At energies near and below the fusion barrier, excitation of the compound nucleus is modest, E$^*$= 30-50 MeV, and consequently light-particle de-excitation of the compound nucleus results in an evaporation residue (ER) with atomic and mass number that are also higher than those of the beam. The ERs can thus be identified by a marked increase in energy deposit
($\Delta$E) due to their increased atomic and mass number relative to the beam.
The segmentation of the anode means fusion events are associated with
discrete locations (and therefore discrete energies) inside the detector.

\begin{figure}[h]
\begin{center}
  \includegraphics[scale=0.500]{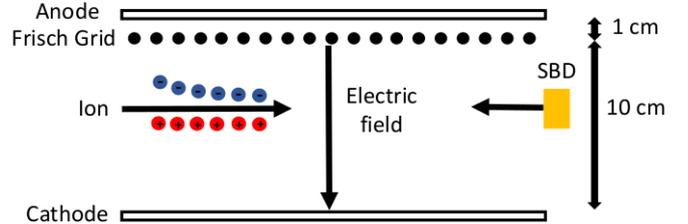}
\caption{
Schematic side view of MuSIC@Indiana. Insertion of the SBD from downstream 
into the active volume is also indicated.
}
\label{fig:MUSICDesign}
\end{center}
\end{figure}

The overall design of MuSIC@Indiana is similar to other 
MuSIC detectors presently in use \cite{Carnelli14, Avila17, Asher21}. The active volume is formed by six printed circuit boards which together constitute a rectangular box. The top and bottom of the box serve as the anode and cathode respectively. Between the anode and cathode is a wire plane (50 $\mu$m diameter Au-W wires on a 1 mm pitch) that acts as a Frisch grid.
A side view of MuSIC@Indiana indicating the anode-to-Frisch grid and
Frisch grid-to-cathode spacings is presented in Fig. \ref{fig:MUSICDesign}. To provide a short collection time of the primary ionization produced by an incident ion, the detector was operated at
a reduced electric field of $\sim$0.7 kV/cm/atm between the cathode and the Frisch grid.
This field yields an electron drift velocity of $\sim$10 cm/$\mu$s in both 
CH$_{4}$ and CF$_{4}$ \cite{Foreman81,Vavra93}.
A significantly higher reduced electric field between the Frisch 
grid and the anode  ($\sim$1.4 kV/cm/atm)
minimizes termination of electrons on the Frisch grid.
Field shaping at the edges of the detector is accomplished using printed circuit boards with 1.613 mm strips and a center-to-center pitch of 3.226 mm.
 A 30 mm diameter hole in the upstream and downstream PCB boards allows the beam to enter and exit the active volume of the detector.
The hole in the downstream PCB also enables the precise 
insertion of a small silicon surface barrier detector (SBD) using a 
linear-motion vacuum feedthrough (Huntington L-2211-6). This ability to insert a SBD precisely into the active volume is critical in the calibration and operation of MuSIC@Indiana.

\begin{figure}[h]
\begin{center}
  \includegraphics[scale=0.240]{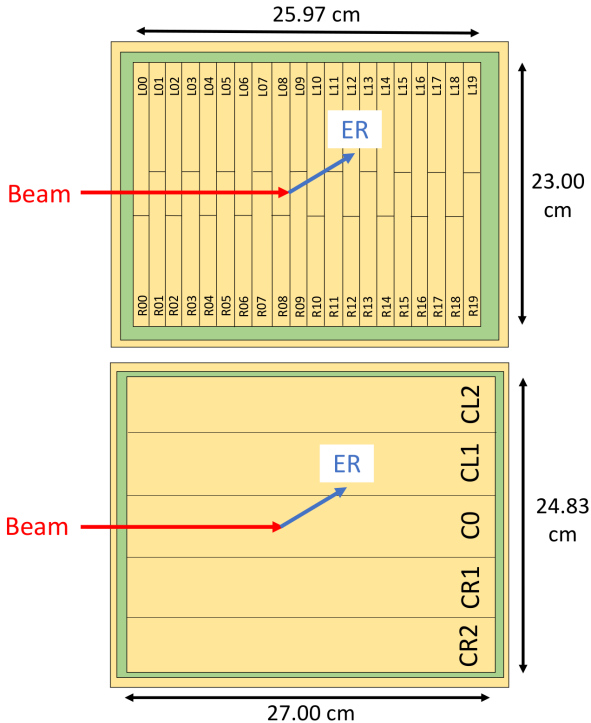}
\caption{
Schematic layout of the MuSIC@Indiana anode (top) and cathode (bottom).  The dimensions given are for the
active areas of the detector.
}
\label{fig:AnodeCathode}
\end{center}
\end{figure}

The dimensions of the active area of MuSIC@Indiana are indicated in Fig. \ref{fig:AnodeCathode}.
The relatively large width of MuSIC@Indiana means 
the measurement of ER energy loss will have high
efficiency even for reactions where ERs reach angles as large as 45$^{\circ}$.
The anode in MuSIC@Indiana is subdivided into 20 distinct 
segments along the beam direction.  Further segmentation transverse to the beam direction provides the left (L0-L19) and 
right (R0-R19) geometry depicted in Fig. \ref{fig:AnodeCathode}.
Each anode segment is 1.219 cm wide with a 0.031 cm inter-strip separation between anodes.
This width for an anode segment along the beam direction was chosen to provide a sufficiently large $\Delta$E signal to yield a good signal-to-noise ratio. When the detector is operated at P = 150 Torr of CH$_4$ gas, an incident $^{18}$O ion with E$_{lab}$ = 50 MeV deposits a $\Delta$E of $\sim$1.5 MeV for an anode. 
 
Anode 0 is used as a "control anode" to reject fusion or scattering events from the beam
on nuclei in the entrance window or gas prior to entering the detector active volume.
For adjacent anode strips, left and right anode strips alternately overlap the 0$^{\circ}$ beam path by 1 cm as indicated in Fig. \ref{fig:AnodeCathode}.
This left-right geometry has been successfully used in other MuSIC detectors \cite{Carnelli14} to distinguish fusion events from two-body 
scattering (discussed further in Section \ref{section:Analysis}).

The cathode is divided into 5 strips which run parallel 
to the beam direction.  These strips are labeled C0, CL1-2, 
and CR1-2 as illustrated in the schematic shown in 
Fig. \ref{fig:AnodeCathode}.  The labels "CL" and "CR" on these strips correspond to the beam-left and beam-right cathode strips respectively. Segmentation of the cathode reduces its capacitance making the capacitance of each cathode strip comparable to an anode segment enabling a fast response for the sensing of the electron motion away from the cathode.

\begin{figure}[h]
  \begin{center}
\includegraphics[scale=0.100]{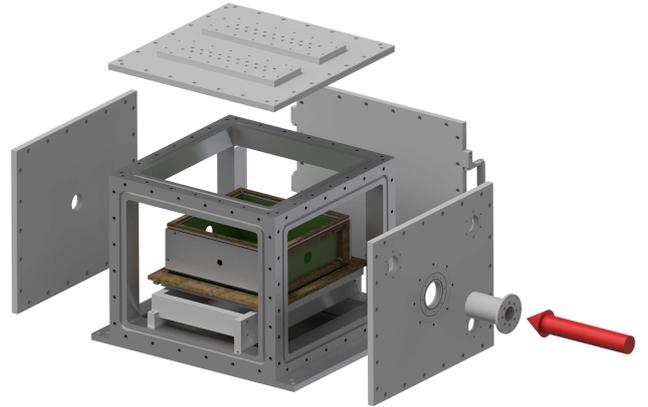}
\caption{
CAD of the chamber that houses the active region of MuSIC@Indiana.
The arrow indicates the direction of the incident beam.
}
\label{fig:Chamber}
\end{center}
\end{figure}

The active detector is housed inside of a 18" (W) x 18" (L) x 15.5" (H) chamber which was machined from
a solid block of aluminum by Indiana University Mechanical Instrument Services resulting in a cube as shown in Fig. \ref{fig:Chamber}. This fabrication approach ensures a clean machined interior surface which is free of welds. The six sides of the cube are sealed by six large flanges with 'O' rings. SMA electrical feedthroughs transport the 40 anode signals through two flanges situated on the top flange. Connected to each of these two flanges is a motherboard housing 20 high-quality charge sensitive amplifiers (CSAs) \cite{Zeptosystems}. Coaxial cables transmit the CSA output to 
analog electronics which process the signals before being 
recorded by the data acquisition.
Signals from the cathode, biasing of the anode and cathode, along with pumping and gas inlet and outlet are provided on the bottom flange. The upstream flange provides a re-entrant window  while the downstream flange provides the means to insert the SBD detector previously described using a linear positioner. The re-entrant window provides separation of the gas volume from the vacuum upstream. During an experiment, this window consists of a 2.6 $\mu$m doubly-aluminized mylar window sealed with an 'O' ring. This window was successfully tested to a pressure of 200 Torr. A thinner window of 1.5 $\mu$m doubly-aluminized mylar exhibited leakage at a pressure of 100 Torr.

As it functions as an active target, maintaining contaminant 
free gas at a stable pressure is critical to the proper operation of MuSIC@Indiana. This was accomplished by using an oil-free gas handling system (GHS). During operation, gas was continuously flowed through the detector via the GHS with the flow controlled by an electronic valve/controller (MKS 0248D-00500RV). The gas flow rate was chosen so that the gas volume of the detector ($\approx$55 L)  was replenished in approximately one hour. Feedback for the solenoid valve was provided by 
monitoring the pressure inside the detector using a MKS 
Model 226 Differential Pressure Transducer.
With this GHS it was possible to maintain a stable pressure in the detector to within 0.1 Torr of the set pressure. The pressure inside MuSIC@Indiana was independently measured using a piezovaccum transducer (Newport 902B) with an accuracy of 0.1 Torr.

First tests on MuSIC@Indiana were carried out using a 
spectroscopy-grade 105 nCi $^{148}$Gd disk source which emits 
a 3.183 MeV $\alpha$ particle.  To ensure that the entire $\alpha$ energy was deposited over a single anode, the detector was operated at a pressure of 400 Torr of CF$_{4}$ gas. The source was then positioned over each anode segment and the energy deposited by the $\alpha$ particle over that  
segment was measured. Under these conditions, the adjacent segments showed no appreciable energy deposit from the $\alpha$ particle.
The results of these bench tests revealed that the inherent resolution of 
each anode is $\sim$100 keV FWHM.

\section{Characterization of MuSIC@Indiana with beam}
\label{section:Characterization}
The fusion excitation function for $^{18}$O+$^{12}$C has 
been well measured \cite{Kovar79,Steinbach14,Eyal76,Heusch82} and therefore provides
a useful reference measurement for the commissioning of MuSIC@Indiana.
To measure this excitation function,
a beam of $^{18}$O$^{6+}$ ions was accelerated to an energy of E$_{lab}$ = 55 MeV by the Notre Dame Nuclear
Science Laboratory's 
10MV Tandem Accelerator. The beam intensity was reduced to an intensity of $\sim$10$^{4}$ particles/s in a contrtolled manner by passing it through slits and a 1/1000 sieve well upstream of the setup. The resulting low-intensity beam was focused onto
MuSIC@Indiana filled with CH$_{4}$ gas at a pressure of 150 Torr. The cathode and anode were biased to voltages of -1500 V and 400 V respectively with the Frisch grid held at ground. The fusion excitation function measured was acquired in just 10 hours.

\begin{figure}[h]
\begin{center}
    \includegraphics[scale=0.400]{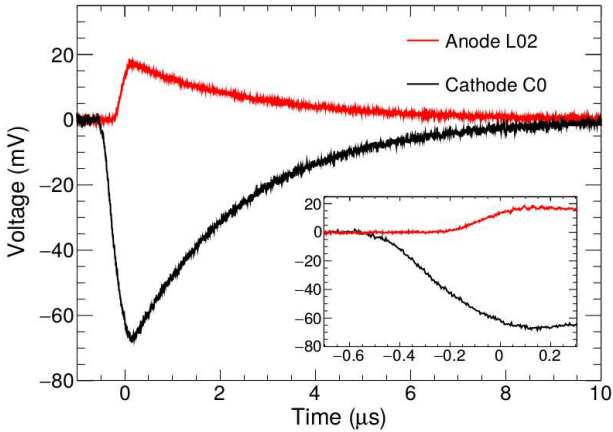}
\caption{
Representative CSA signals from single MuSIC@Indiana anode and cathode strips.
The inset focuses on the rising edge of the same signals.
}
\label{fig:DetectorSignals}
\end{center}
\end{figure}

Representative anode and cathode signals from a putative fusion 
event, processed by the high quality charge-sensitive amplifiers \cite{Zeptosystems}, are presented in Fig. \ref{fig:DetectorSignals}. 
These CSAs yield $\approx$9 mV amplitude signal for a 3.183 MeV $\alpha$
particle in 400 Torr of CF$_{4}$.
Collection of electrons by the anode together with inversion by the CSA
determines the polarity of the anode signal. The risetime of this signal is
approximately 100 ns as evident in the inset of Fig.~\ref{fig:DetectorSignals}, consistent
with the drift velocity of $\sim$10 cm/$\mu$s and the Frisch grid to anode spacing of 1 cm. Examination
of the cathode signal reveals  a much larger amplitude 
which can be understood by noting that the cathode integrates the entire energy along the beam direction 
while the anode only collects a small 
portion of the particle's total ionization. The risetime of the cathode signal 
($\sim$500 ns) is observed to be slower than the anode. This difference is due to the 
larger cathode-to-Frisch grid spacing as compared to the one for the anode-to-Frisch grid. It is also 
observed that the cathode signal precedes the anode 
signal by 400 ns. The delay of the anode relative to the cathode is
due to the shielding of the anode from electron motion until they have passed the 
Frisch grid. It should be noted that the signal observed for the cathode is not due to the motion of the cations but 
due to the motion of the electrons away from the cathode.
The fall time of the CSA signals is only $\sim$8 $\mu$s
which allows successful operation of MuSIC@Indiana at a rate up to 1 x 10$^5$ particles/s. 

The CSA signals from the detector are processed through standard shaping amplifiers and peak
sensing digitizers (CAEN V785 ADC) before being acquired by the VME data acquisition 
system (DAQ) and recorded on
the computer. The DAQ was triggered using signals from the segmented cathode. Each cathode's CSA signal was processed by a timing filter amplifier (TFA) and shaping amplifier. The TFA signals were summed, discriminated and the resulting logical signal was used to gate the ADCs as well as trigger the data acquisition system.

\begin{figure}[h]
\begin{center}
  \includegraphics[scale=0.440]{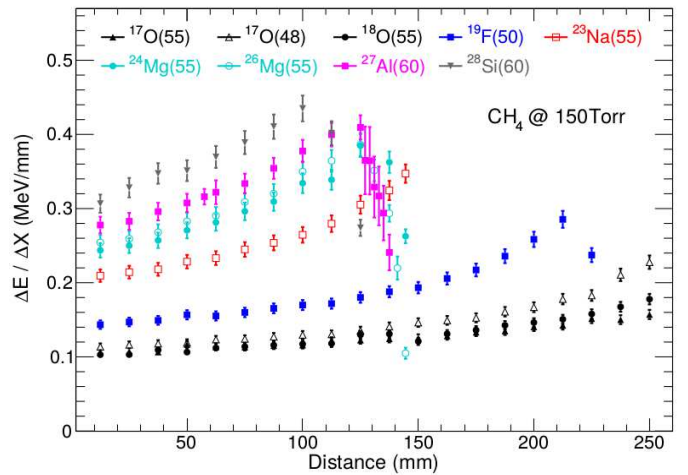}
\caption{
Measured energy loss for several ions including potential 
residues and the beam.  Listed along each isotope is the
 incident ion's energy in MeV.
}
\label{fig:EnergyLoss}
\end{center}
\end{figure}

Measuring the fusion excitation function requires knowledge of the incident energy across each anode. To measure this energy, a surface barrier detector was inserted from downstream into the active volume of the detector. Use of a precision linear positioner allowed the SBD to be positioned at the front and back of each anode with an accuracy of 0.5 mm. At these positions the energy of the beam was recorded at low beam intensity. The result of this measurement is shown in Fig.~\ref{fig:EnergyLoss}. In addition, the energy loss curves for several heavier products was also measured by impinging low-intensity beams of $^{19}$F, $^{23}$Na, $^{24}$Mg, $^{26}$Mg, $^{27}$Al, and $^{28}$Si with E$_{lab}$ = 50-60 MeV on the detector. The resulting energy loss curves are presented in Fig.~\ref{fig:EnergyLoss}. These measurements characterized the response of MuSIC@Indiana making the use of energy loss programs such as SRIM \cite{SRIM} unnecessary. It has been established that energy loss programs have uncertainties of approximately 10\% \cite{Carnelli15, Avila17}.

\section{Simple data analysis for extraction of the $^{18}$O+$^{12}$C fusion excitation function}
\label{section:Analysis}

\begin{figure}[h]
\begin{center}
  \includegraphics[scale=0.400]{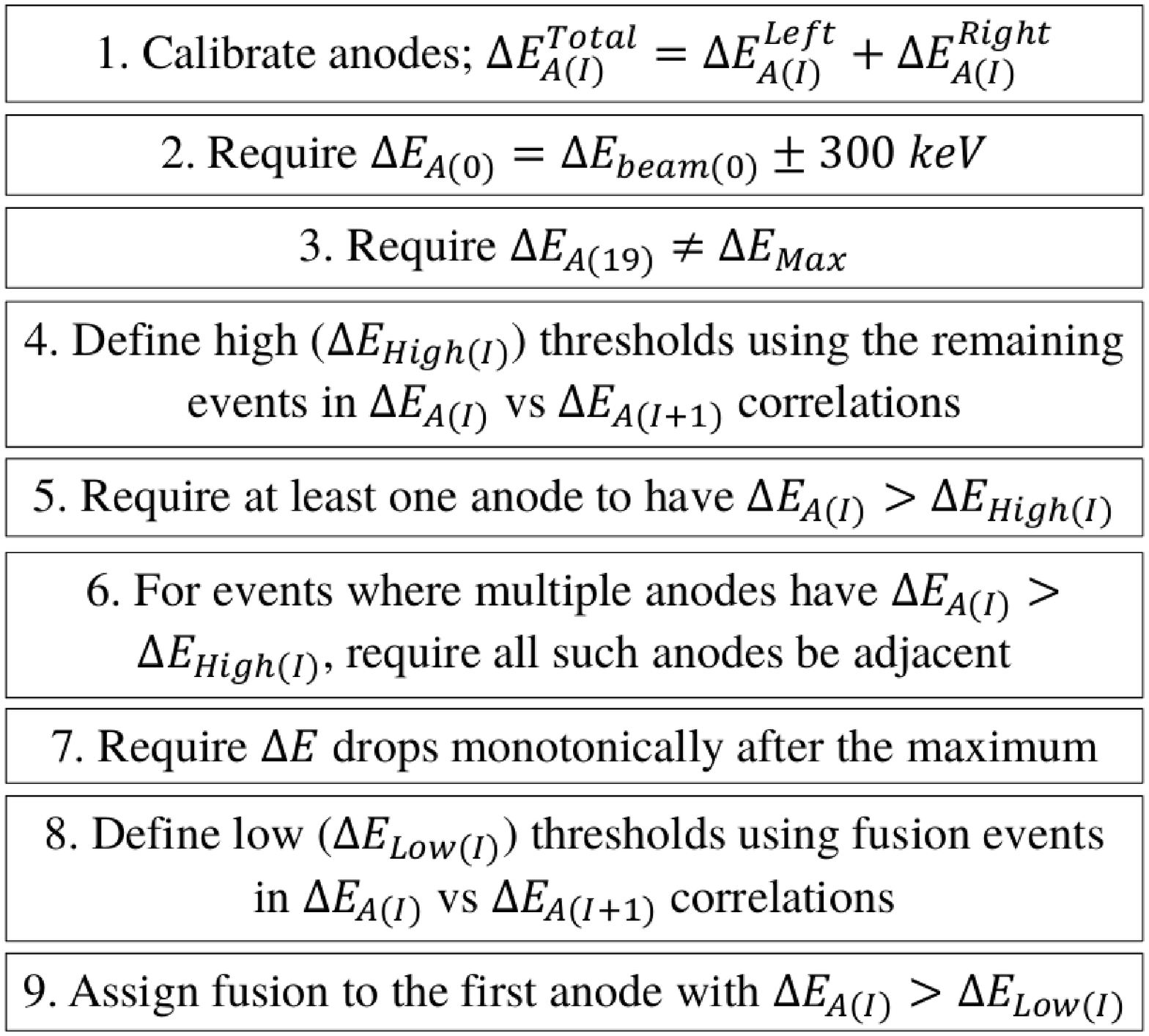}
\caption{
MuSIC@Indiana analysis logic flowchart.
}
\label{fig:AnalysisFlowchart}
\end{center}
\end{figure}

Schematically illustrated in Fig. \ref{fig:AnalysisFlowchart} is the sequence for analyzing data from MuSIC@Indiana.
The initial step in the analysis of MuSIC data involves the calibration of the left and 
right anode segments using the energy loss of the beam as measured by the SBD. 
Once the left and right segments have been calibrated, 
the two sides can be summed to calculate the
total energy loss in an anode.
For all subsequent steps in the analysis only the summed anode energy loss is used.

The second step in the analysis is to require the incident ion have the $\Delta$E of the beam ($\pm$ 300 keV) in anode 0.
This step is critical to accurately measuring the excitation function as it eliminates
events where fusion occurs in either the window or gas upstream of the active area of MuSIC@Indiana. It also removes any beam pileup events.

The next step in the MuSIC@Indiana analysis requires that the anode of maximum
energy loss is not anode 19.  Similar to anode 0, anode 19 is used as a control anode.
Requiring $\Delta$E$_{A(19)}$ ${\neq}$ $\Delta$E$_{Max}$ 
removes events (both fusion and scattering) 
which occur in anode 19.

\begin{figure}[ht]
\begin{center}
  \includegraphics[scale=0.500]{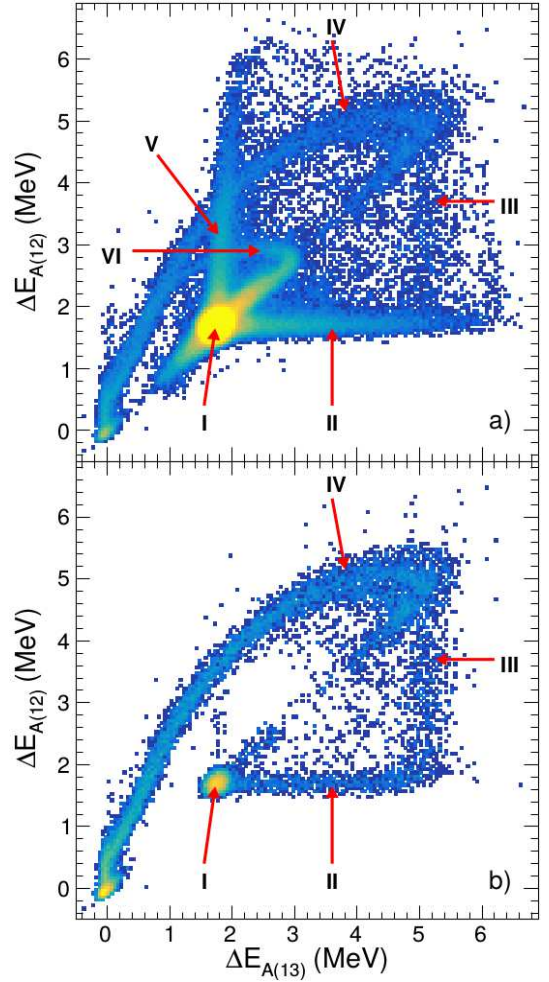}
\caption{
${\Delta}$E$_{A(12)}$ vs ${\Delta}$E$_{A(13)}$ correlations.  
The upper plot shows the correlation after
the first three steps of the analysis.  
The bottom plot shows the correlation only for events which 
were identified as ERs.  Features of the correlations are identified by numerals
and are explained in the text.
}
\label{fig:AnodeCorrelation2D}
\end{center}
\end{figure}

After these first two requirements have been implemented, 
the correlation between the deposited energy in an anode and the energy deposit in the subsequent anode is examined.
A representative correlation is shown in Fig. \ref{fig:AnodeCorrelation2D}a for anodes 12 and 13.
Several features appear in the correlation each of which was 
identified by examining plots of $\Delta$E vs anode number
(called traces) associated with each feature.

\paragraph{Feature I} The most prominent feature in the spectrum is
the bright spot at $\Delta$E$_{A(12)}$ = 1.8 MeV and $\Delta$E$_{A(13)}$ = 1.8 MeV
which corresponds to events which are beam in both anode 12 and anode 13.

\paragraph{Feature II} The second feature is a horizontal band which extends from the 
beam peak out to $\Delta$E$_{A(13)}$$\approx$6 MeV.
This feature corresponds to events where fusion occurs in anode 13. The fusion product with its larger atomic and mass number than the beam has a larger specific ionization and consequently a larger energy deposit.

\paragraph{Feature III} Extending vertically from $\Delta$E$_{A(12)}$ = 1.8 MeV 
and $\Delta$E$_{A(13)}$ = 5.2 MeV is a faint line which 
corresponds to events where fusion occurred in anode 12.

\paragraph{Feature IV} Starting from $\Delta$E$_{A(12)}$ = 
5.2 MeV and $\Delta$E$_{A(13)}$ = 5.2 MeV 
and extending as a tail to lower $\Delta$E are events 
where fusion occurred in anodes prior to anode 12.
This locus terminates at a distinct peak at $\Delta$E$_{A(12)}$ = 0 MeV and $\Delta$E$_{A(13)}$ = 0 MeV.
This peak is associated with events in which fusion occurred much earlier in the detector and the 
ER has already ranged out in the detector gas prior to anode 12.

\paragraph{Feature V} The near-vertical band extending from the beam peak corresponds to two-body events which are subsequently eliminated in the analysis.

\begin{figure}[h]
  \begin{center}
    \includegraphics[scale=0.400]{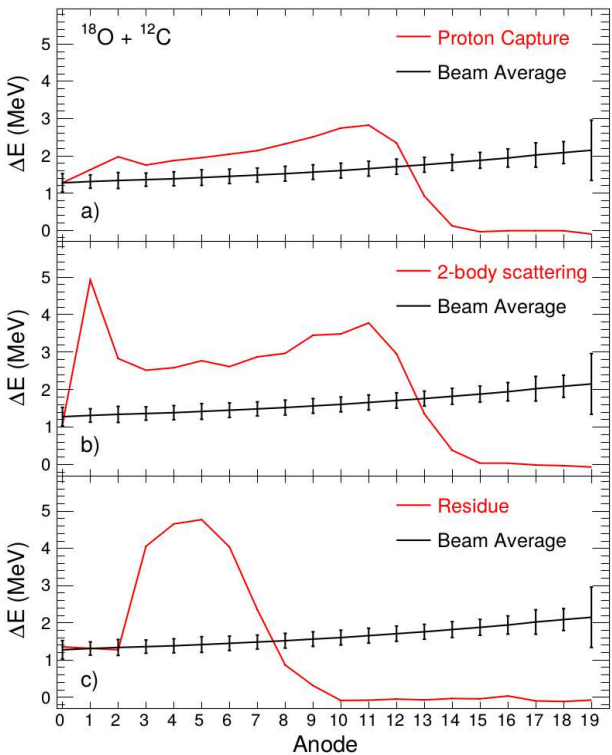}
\caption{
Experimental MuSIC@Indiana traces. Panel a) shows the trace for a proton capture event.  Panel b) shows the trace for a two-body scattering event.
Panel c) shows the trace for a residue from fusion occurring in anode 3. For all panels, the average beam trace is shown as the black line. 
The error bars on the average beam trace represent the FWHM of the
beam $\Delta$E distribution in that anode.
}
\label{fig:Traces}
\end{center}
\end{figure}

\paragraph{Feature VI} Extending diagonally from the beam peak up 
to $\Delta$E$_{A(12)}$ = 2.8 MeV and $\Delta$E$_{A(13)}$ = 2.8 MeV, 
and then turning with a tail back down to $\Delta$E$_{A(12)}$ = 0 MeV and $\Delta$E$_{A(13)}$ = 0 MeV are proton capture events resulting
from the fusion of beam on hydrogen in the CH$_{4}$ detector gas.
Proton capture of the $^{18}$O beam results in $^{19}$F which exhibits a larger specific ionization than the beam but less than that of the ERs. These
proton capture events are characterized by $\Delta$E values that are higher than the beam 
for several consecutive anodes before dropping to $\Delta$E = 0 MeV at the end of the detector.
A representative proton capture event is presented in Fig. \ref{fig:Traces}a

Correlations like the one shown in Fig. \ref{fig:AnodeCorrelation2D}a are used 
in the analysis to establish the quantity ${\Delta}$E$_{High(I)}$ for each anode.  For example,
Fig. \ref{fig:AnodeCorrelation2D}a was used to set ${\Delta}$E$_{High(12)}$ = 3.2 MeV, a measure 
of the maximum energy deposit associated with proton capture for that anode. Requirement that at least one anode
has ${\Delta}$E$_{A(I)}$ ${>}$ ${\Delta}$E$_{High(I)}$ eliminates proton capture events from the data leaving putative fusion events.

After removing the proton capture events, the analysis requires that if
multiple anodes have ${\Delta}$E$_{A(I)}$ ${>}$ ${\Delta}$E$_{High(I)}$, 
those anodes must be adjacent. This requirement rejects the majority of two-body scattering events.
A representative two-body scattering event is shown in Fig. \ref{fig:Traces}b. Two-body scattering events are characterized by two particles with different specific ionization and consequently two ranges. This
behavior is clearly evident in Fig. \ref{fig:Traces}b where one particle has a range of 1-2 anodes and the other has a range of approximately 11 anodes.
Observation of two Bragg 
peaks in the trace is a clear indication of two particles in a single event.

Not all two-body scattering events are eliminated through the previous analysis step.
To remove the remaining scattering events, the analysis requires that the ${\Delta}$E drops monotonically after the anode of maximum ${\Delta}$E.
This requirement specifically eliminates scattering events where the low specific ionization (beam-like) particle
has a long track and does not pass the high threshold later in the detector.
This analysis is distinct from previous MuSIC analyses 
which required the use of the detector's
left/right anode structure to distinguish two-body 
events from fusion events \cite{Carnelli15}.
While events containing two particles can be eliminated by using the 
left/right information, the analysis steps described in this paper provide a 
simple and equally effective way of removing two-body scattering events from the data.

All remaining events are assigned as fusion events. A representative trace of a fusion event is presented in Fig. \ref{fig:Traces}c. Prior to anode 2 the $\Delta$E observed is consistent with that of beam. At anode 2 the $\Delta$E increases markedly reaching a maximum at anode 5 whereupon it decreases monotonically until anode 10. No additional energy is observed at subsequent anodes.
Using the fusion events selected in this manner, correlations like the
one shown in Fig. \ref{fig:AnodeCorrelation2D}b are made.
After all of the analysis steps have been followed, Features V and VI as well as the 
scattering events in Feature II have been removed.  The residues appearing in Features 
II, III, and IV are now clear.  The remaining events in Feature I of 
Fig. \ref{fig:AnodeCorrelation2D}b correspond to fusion events which 
happen in anodes after anode 13. Using this correlation a low threshold, ${\Delta}$E$_{Low(I)}$, is established for each anode.
This threshold is set just above Feature I and is used to assign the anode of fusion.
For example, Fig. \ref{fig:AnodeCorrelation2D}b was used to set 
${\Delta}$E$_{Low(12)}$ = 2.2 MeV.
The anode of fusion is assigned to the first anode with 
${\Delta}$E$_{A(I)}$ ${>}$ ${\Delta}$E$_{Low(I)}$.

\begin{figure}
\begin{center}
  \includegraphics[scale=0.400]{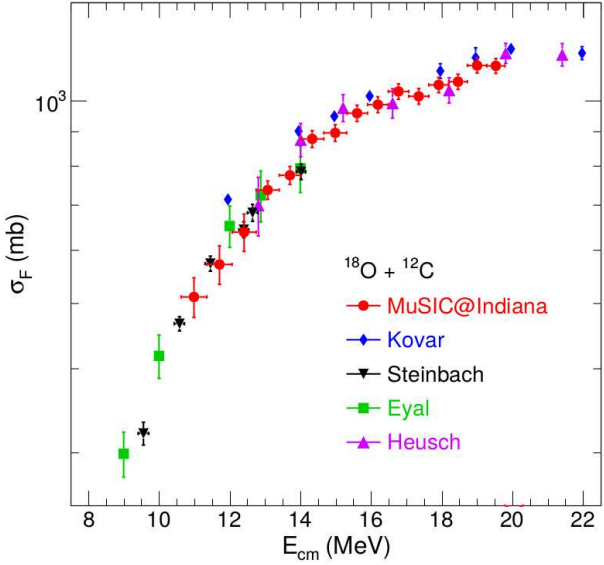}
\caption{
Fusion excitation function of $^{18}$O+$^{12}$C.
The literature datasets are Kovar \cite{Kovar79},
Steinbach \cite{Steinbach14}, Eyal \cite{Eyal76}, and Heusch \cite{Heusch82}.
}
\label{fig:ExcFunc}
\end{center}
\end{figure}

Once the occurrence of fusion has been identified and assigned to the appropriate
anode, the cross-section can be calculated using: $\sigma = N_{ER}/(\epsilon*I*t)$, 
where N$_{ER}$ is the number of ERs in an anode, $\epsilon$ is the detector efficiency
of an anode, I is the number of incident beam particles, 
and t is the target thickness as defined by the anode width and the gas pressure.

Use of anodes 0 and 19 as control anodes prohibits using them in the measurement of the excitation function. Moreover, the most downstream anodes are less than 100\% efficient. An anode can be considered 100\% efficient if there are a sufficient number of anodes to observe the peak in the corresponding trace, which requires 4-5 anodes. Consequently, the fusion cross-section is measurable for anodes 1 - 15 with 100\% efficiency. Given this intrinsic efficiency, no efficiency correction is necessary to extract the fusion cross-section. In the present measurement this enabled us to measure the fusion cross-section for 11 MeV $<$ E$_{cm}$ $<$ 20 MeV.

In order to assign an energy to the cross-section associated with
a particular anode, the SBD measurement of the beam at the front
and back of each anode was used.
Each datapoint in the excitation function is initially assigned the energy in the middle of the associated anode (calculated as the average of the energies at the front and the back of the anode). In reality the energy average of the anode
is weighted toward higher energy where the cross-section is higher.  
To correct for this, the initial excitation function is
parameterized using a Wong formalism \cite{Wong73}.  Each datapoint is
then segmented into 20 equally-spaced slices in energy.
The new energy is calculated as the weighted average of the energy of the slices using the Wong-calculated cross-section: 
\[E' = \frac{\sum[{\sigma}_{Wong}(slice)*E(slice)]}{\sum{\sigma}_{Wong}(slice)}\]

This process is repeated until the energy converges.
Horizontal error bars represent the difference between the assigned anode energy and the energy at the front/back of the anode.
The size of the vertical error bars are calculated from the experimental statistics.

The measured $^{18}$O+$^{12}$C excitation function from this work is displayed in Fig. \ref{fig:ExcFunc}. It is observed to be in good agreement with the previously reported cross-section in the literature.
Below E$_{cm}$ = 14 MeV the present work matches the 
Eyal \cite{Eyal76} and Steinbach \cite{Steinbach14} measurements even
to the extent of interpolating between the published points in those datasets.
In this same region there are two datapoints from 
Kovar \cite{Kovar79} and one datapoint from Heusch \cite{Heusch82}
which are high relative to Eyal, Steinbach, and the present work.
This result suggests that Kovar and Heusch may provide a systematically high measurement of the cross-section.
This trend continues above E$_{cm}$ = 14 MeV with the 
present work's cross-sections below all Kovar datapoints.
The present work's excitation function overlaps two points from Heusch
but these points are also low relative to all other Heusch points. It should be appreciated that the prior measurements of the fusion cross-section were thin-target 
measurements with limited angular coverage. Extraction of the fusion cross-section required integration of the angle and energy distributions for the individual ERs which introduces uncertainties into the total extracted fusion cross-section. Use of a MuSIC detector provides a direct integrated measure of the fusion cross-section.

\section{Conclusions}
\label{section:Conclusions}
MuSIC detectors with their direct measurement of the angle-integrated fusion cross-section and ability to simultaneously measure multiple points on an excitation function are a powerful tool  for radioactive beam experiments, particularly for intensities below 10$^{4}$-10$^{5}$ particles/s.
MuSIC@Indiana is differentiated from other MuSIC detectors in its 
ability to precisely insert an SBD into the detector active volume.
The SBD enables the accurate measurement of the beam 
energy at each anode allowing calibration of
MuSIC@Indiana and eliminates the uncertainties associated with energy loss programs in constructing the fusion excitation 
function.  To characterize the response of the detector to evaporation residues the energy loss of heavier ions in the CH$_4$ gas was also measured. An analysis procedure was developed that provided a simple means of discriminating fusion events from proton capture and two-body scattering events. The effectiveness of this analysis was demonstrated by the good agreement between the extracted excitation function and 
previously reported cross-sections.
The quality of the $^{18}$O+$^{12}$C  measurement over a short time interval demonstrates that
MuSIC@Indiana is an effective tool for accurate measurement of fusion with low-intensity radioactive beams.

\section{Acknowledgements}
The authors would like to thank Indiana University Mechanical Instrument Services
and Electronic Instrument Services in the Department of Chemistry for their help in constructing the MuSIC@Indiana
detector.  This work was supported by US Department of Energy under Grant No.
DE-FG02-88ER-40404. Research sponsored by NSF Grant No. PHY-2011890 and by the University of Notre Dame.


\end{document}